\documentclass{revtex4}
\usepackage{amsmath}
\usepackage{amssymb}
\usepackage{graphicx}

\begin{document}

\title{Two interacting GL-equations in High-T$_c$ superconductivity and quantum chromodynamics}
\author{Vladimir Dzhunushaliev
\footnote{Senior Associate of the Abdus Salam ICTP}}
\email{dzhun@krsu.edu.kg}
\affiliation{Dept. Phys. and Microel.
Engineer., Kyrgyz-Russian Slavic University, Bishkek, Kievskaya Str.
44, 720021, Kyrgyz Republic}

\begin{abstract}
The possible connection between High-T$_c$ superconductivity and quantum chromodynamics is considered that is based on two interacting Ginzburg-Landau equations. For High-T$_c$ superconductivity these two equations describe Cooper electrons interacting with different kind of quasi particles (phonons, magnons, excitons and so on). The interaction term describes a  possible interaction between different kind of quasi particles. For quantum chromodynamics the  equations describe two kinds of gauge condensates. The condensates describe a gauge potential from a subalgebra of the SU(3) gauge group and the corresponding coset. Regular solutions are found which describe the situation where one field, $\psi_1$, is pushed out by  another, $\psi_2$.
%
\end{abstract}


\maketitle

\section{Introduction}

The idea that superconductivity and quantum chromodyanmics have common features has long history. Probably the first time this idea was considered in Ref.~\cite{thooft} where it was suggested that the confinement of quarks into hadrons may happen due to a condensation of a special gluonic configurations called Abelian monopoles. In this approach -- referred to as the dual superconductivity scenario -- a condensate of the monopoles breaks spontaneously an internal $U(1)$ gauge symmetry. According to the dual superconductor idea, the breaking of the dual symmetry gives rise naturally to the dual Meissner effect, which insures a formation of a QCD flux tube, which in turn leads to the confinement the quarks into hadronic bound states. In this paper we want to explore a similar idea in both directions: from quantum chromodynamics to High-T$_c$ superconductivity and vice versa. 
\par
In High-T$_c$ superconductivity there is an opinion that a few kind of quasi particles (phonos, magnons, excitons and so on) may play the role of carrying the interaction between electrons in High-T$_c$ superconductor. It is similar to Yang-Mills gauge fields where the gauge potential $A^B_\mu \in SU(n)$ has several components $B = 1,2, \ldots , n$. In non-Abelian gauge theories also it is important that the different $A^B_\mu$ and $A^C_\mu, (B \neq C)$ interact between themselves. We transfer this observation to High-T$_c$ superconductivity and suppose that for every kind of Cooper pair (where the interaction is carried with the definite quasi particles: phonons, magnons and so on) we can write the Ginzburg-Landau (GL)-equation and all GL-equations have interaction terms describing the interaction between different kinds of quasi particles. 
\par 
In quantum chromodynamics we use the idea from superconductivity that the Cooper electrons (which have an interaction with phonons) can be described (in the first phenomenological approximation) with the GL-equation. In quantum chromodynamics this idea takes the form: Every two gluons have an interaction between themselves via gluons. Therefore we assume that such interacting gluons can be described using the GL-equation. But in this case gluons do not form pairs like Cooper pairs and the wave function $\psi$ in the GL-equation will describe not pairs of gluons but a condensate made of \emph{all} gluons. Another feature of this phenomenological approach in quantum chromodynamics is that we assume that there are two kind of condensate: the first one describes correlated gluons (condensate $\psi_1$) belonging to an $SU(2) \subset SU(3)$ subgroup and the second one describes correlated gluons (condensate $\psi_2$) belonging to the coset $SU(3)/SU(2)$. 
\par 
The main goal of the idea presented here is to show that: (a) a High-T$_c$ superconductor may have the space-separated regions where the interaction between Cooper electrons is carried out by different kind of quantum (either phonons, or magnons, or excitons or and so on); (b) the similar situation may happen in quantum chromodynamics where the regions filled with different kind of a nonperturbative gauge condensate may exist. 

\section{Two interacting GL-equations in High-T$_c$ superconductivity}

In High-T$_c$ superconductivity there exist several candidates for the role of pseudoparticles which carry the interaction between electrons: phonons, magnons, excitons and so on. Some researchers argue that patterns of charges, currents, waves of magnetism, or phonons play a crucial role for understanding High-T$_c$ superconductivity. Pairing might even require all of these things in combination. In this paper we assume that there is an interaction between phonons, magnons and so on. From this point of view we consider the following approximate phenomenological model: for every kind of interaction (phonons, magnons and so on) we can write the GL-equation. In the consequence of such an assumed interaction between phonons, magnons and so on we have to have interaction terms in every GL-equation. 
\par
For the simplicity we will consider here only two connected GL-equations corresponding to two kind of interactions between electrons in High-T$_c$ superconductors. Let us at first consider the possible form of the interaction term. For simplicity we will assume that the term is 
\begin{equation}
	V_{int} = \alpha \left| \psi_1 \right|^2 \left| \psi_2 \right|^2
\label{1-10}
\end{equation}
where $\alpha$ is a constant; $\psi_{1,2}$ are wave functions describing Cooper pairs where the electrons interact either via phonons or, magnons or another quanta. We suppose also that $\alpha > 0$ which means that the interaction is repulsive. As we see later this leads to the displacement of one condensate by another. This term can be considered as the first nonlinear term in the Taylor expansion of any potential $V\left( \left| \psi_1 \right|^2, \left| \psi_2 \right|^2 \right)$. 
\par
After some redefinitions of coefficients of two interacting GL-equations they can be written in the following way
\begin{eqnarray}
	- \Delta \psi_1 + \left[ 
		\left| \psi_2 \right|^2 + \lambda_1 \left(
			\left| \psi_1 \right|^2 - \frac{E_1}{\lambda_1}
		\right)
	\right] \psi_1 &=& 0, 
\label{1-20} \\ 
	- \Delta \psi_2 + \left[ 
		\left| \psi_1 \right|^2 + \lambda_2 \left(
			\left| \psi_2 \right|^2 - \frac{E_2}{\lambda_2}
		\right)
	\right] \psi_2 &=& 0 
\label{1-30}
\end{eqnarray}
where $\lambda_{1,2}$ constants, $E_{1,2}$ parameters. The most interesting is that they have \emph{regular} solutions which are missing in a single GL-equation. In the author's opinion these regular solutions describe the following situation: there exists a ground state filled with $\psi_1$ or $\psi_2$, then the solution are plane, cylindrically or spherically symmetric patterns in the ground state. In other words $\psi_1$ and $\psi_2$ repel each other.
\par 
The potential energy of the interacting wave functions $\psi_{1,2}$ is 
\begin{equation}
	V\left( \psi_{1,2} \right) = \frac{\lambda_1}{4} \left(
		\left| \psi_1 \right|^2 - \frac{E_1}{\lambda_1}
	\right)^2 + \frac{\lambda_2}{4} \left(
		\left| \psi_2 \right|^2 - \frac{E_2}{\lambda_2}
	\right)^2 + 
	\frac{1}{2} \left| \psi_1 \right|^2 \left| \psi_2 \right|^2 
\label{1-40}
\end{equation}
Use of two fields $\psi_{1,2}$ ensures presence of two global minima of the potential  \eqref{1-40} at $\psi_1~=~0, \psi_2~=~\sqrt{E_2/\lambda_2}$ and two local minima at 
$\psi_1~=~\sqrt{E_1/\lambda_1}, \psi_2~=~0$ for values of the parameters $\lambda_1, \lambda_2$ used in the paper. The conditions for existence of the local minima are: 
$\lambda_1>0, E_1/\lambda_1 > E_2/2$, and for the global minima: 
$\lambda_2>0, E_2/\lambda_2 > E_1/2$. The presence of these minima has allowed to find solutions localized on the plane, tube and ball when the solutions have tended asymptotically to one of the local minima.

\section{Two interacting GL-equations in quantum chromodynamics}

In Ref's~\cite{Chernodub:2005jh} \cite{Dzhunushaliev:2000ma} the idea is considered that there is a connection between condensed matter physics and quantum chromodynamics. In Ref.~\cite{Chernodub:2005jh} it is shown that using a spin--charge separation of the gluon field in the Landau gauge the $SU(2)$ Yang-Mills theory in the low-temperature phase
can be considered as a nematic liquid crystal. In Ref's~\cite{Chernodub:2005jh}, \cite{Niemi:2005qs} and \cite{Faddeev:2006sw} the idea of slave-boson decomposition is transferred into quantum chromodynamics which gives us  the so called spin-charge separation for the gauge fields. In Ref.~\cite{Dzhunushaliev:2000ma} it is supposed that phonons have a strong interaction between themselves similar to the strong interaction between gluons in quantum chromodynamics. In quantum chromodynamics such strong interaction leads to a flux tube stretched between quarks. In consequence of this similarity it is supposed that the phonons in a superconductor are confined in a tube similar to the flux tube in quantum chromodynamics. Here we would like to investigate similar connections applying to quantum chromodynamics. 
\par
In a superconductor the Cooper electrons are paired as a consequence of phonon interactions and this picture on the phenomenological level can be considered by using the GL-equation. In quantum chromodynamics any two gluons have an interaction between themselves. One distinction with the superconductivity is that the interaction between quantum (two gluons) is carried by the same quantum (gluons). Nevertheless we assume that (in the first approximation) the GL-equation can be applied for the description of gluons. In fact it will be a non-perturbative, phenomenological picture of gauge fields in quantum chromodynamics. 
\par 
We want to make one distinction in quantum chromodynamics in comparison with the superconductivity. We suppose that the SU(3) gauge fields in quantum chromodynamics are divided into two different fields: the first one is in the subalgebra $SU(2) \subset SU(3)$ and the second one is in the coset $SU(3)/SU(2)$. It means that we have two GL-equations. As SU(3) Yang-Mills Lagrangian has the term describing the interaction between $A^a_\mu \in SU(2)$ and $A^m_\mu \in SU(3)/SU(2)$ fields we should have an interaction term in both GL-equations. 
\par 
Thus we propose to apply two interacting GL-equations \eqref{1-20}-\eqref{1-30} for an approximate description of nonperturbative quantized SU(3) gauge field: $\psi_1$ describes a condensate of nonperturbative quantized $A^a_\mu \in SU(2), a=1,2,3$ gauge field and $\psi_2$ described a condensate of $A^m_\mu \in SU(3)/SU(2), m=4,5,6,7,8$ gauge field. The term 
$\left| \psi_1 \right|^2 \left| \psi_2 \right|^2$ in both GL-equations represents the term 
$\left( f^{Cab} A^a_\mu A^{b \mu} \right) \left( f^{Cmn} A^m_\nu A^{n \nu} \right), C=1, \cdots , 8$ in the SU(3) Yang-Mills Lagrangian. 
\par 
In the section \ref{patterns} we will present numerical solutions which show that in quantum chromodynamics there may exist a situation as follows: The ground state is the condensate $\psi_1$ (or $\psi_2$). In this state there exist patterns where the condensate $\psi_2$ ( or $\psi_1$) pushes out $\psi_1$ (or $\psi_2$). 

\section{Patterns}
\label{patterns}

In this section we would like to present the solutions of two interacting GL-equations with planar, cylindrical and spherical symmetries. 

\subsection{Plane pattern}
\label{plane}

In this case Eq's \eqref{1-20} \eqref{1-30} have the following form 
\begin{eqnarray}
	- \frac{d^2 \psi_1}{dx^2} + \left( 
		\psi_2^2 + \lambda_1 \psi_1^2 
		\right) \psi_1 &=& E_1 \psi_1, 
\label{2-10} \\ 
	- \frac{d^2 \psi_2}{dx^2} + \left( 
		\psi_1^2 + \lambda_2 \psi_2^2 
		\right) \psi_2 &=& E_2 \psi_2 
\label{2-20}
\end{eqnarray}
here we are searching for $\psi_{1,2}$ as real functions; the coordinate $x$ is the Cartesian coordinate that is transversal to the planar pattern. 
\par 
The regular solutions of the system of nonlinear differential equations \eqref{2-10}-\eqref{2-20} could exist only for some values of the self-coupling constants $\lambda_{1,2}$ and the energies $E_{1,2}$. The further task consists in a search of such parameters $E_{1,2}$ which give regular solutions. In this sense the problem reduces to a search of \emph{eigenvalues} of the parameters $E_{1,2}$  and corresponding \emph{eigenfunctions} $\psi_{1,2}$ for two GL-equations \eqref{2-10}-\eqref{2-20}.
\par 
The technique of solution of systems similar to \eqref{2-10}-\eqref{2-20} is described in Ref.~\cite{dzh-step} in details. The essence of this procedure is the following: on the first step one solve the equation \eqref{2-10} with some arbitrary selected function $\psi_2$ looking for a regular solution existing only at some value of the parameter $E_1$. Then this solution for the function $\psi_1$ insert into the equation \eqref{2-20} and one search for a value of the parameter $E_2$ yielding a regular solution. This procedure reiterate several times (six  usually enough) for obtaining of acceptable convergence of values of the parameters $E_{1,2}$.
\par
The described procedure of a search of solutions of the system \eqref{2-10}-\eqref{2-20}, also known as the shooting method, allows to find rather fast values of the parameters $E_{1,2}$ at which regular solutions exist.
\par
\begin{figure}[h]
\begin{minipage}[t]{.49\linewidth}
  \begin{center}
  \includegraphics[width=9cm]{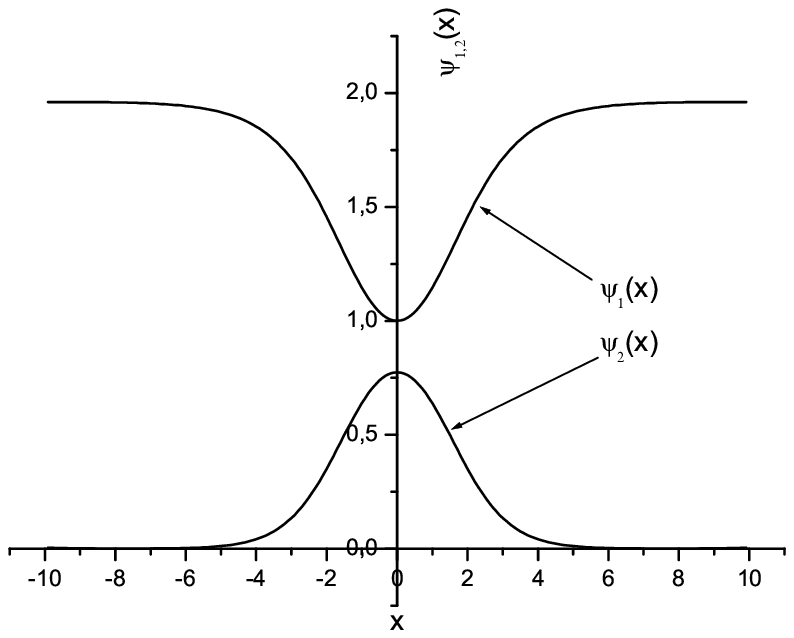}
  \caption{The wave functions $\psi_{1,2}$ for a sandwich-type defect, 
  $\lambda_1 = 0.1, \lambda_2 = 1$.}
  \label{planar}
  \end{center}
\end{minipage}\hfill
\begin{minipage}[t]{.49\linewidth}
  \begin{center}
  \includegraphics[width=9cm]{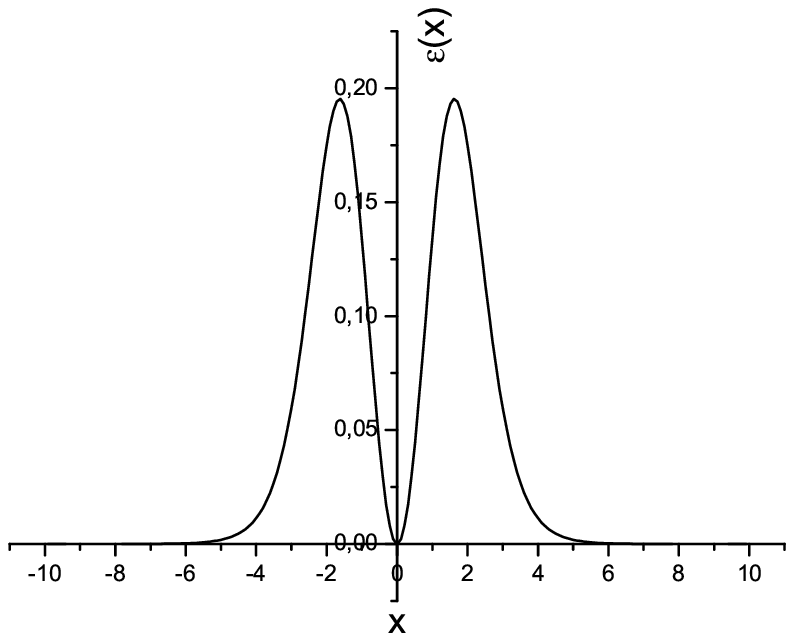}
  \caption{The energy density for for a sandwich-type defect.}
  \label{planar_energy}
  \end{center}
\end{minipage}\hfill
\end{figure}

The boundary conditions are choosing with account of $\mathbb{Z}_2$ symmetry in the following form:
\begin{alignat}{2}
\label{ini1}
	\psi_1(0)	& = 1,				& \qquad \psi_1^\prime(0)	&=0, 
\nonumber \\
	\psi_2(0)	&=\sqrt{0.6},	& \qquad \psi_2^\prime(0)	&=0 
\end{alignat}
Then, using the above procedure for obtaining of solutions of the system  \eqref{2-10}-\eqref{2-20}, we have the results presented in Fig.~\eqref{planar}-\eqref{planar_energy}. These results are obtained for the energies
$E_1 \approx 0.385289$ and $E_2\approx 1.97824$. As one can see from Fig.~\eqref{planar}, $\psi_1 \rightarrow \sqrt{E_1/\lambda_1}$ and $\psi_2 \rightarrow 0$ as 
$x \rightarrow \pm\infty$. It corresponds to asymptotic transition of the solutions to
the local minimum of the potential \eqref{1-40}. 
Let us estimate an asymptotic behaviour of the solutions. For this purpose we will seek for  solutions of the equations \eqref{2-10}-\eqref{2-20} in the form:
\begin{equation}
\label{asymptotic}
	\psi_1 = \frac{E_1}{\lambda_1} - \delta \psi_1, \quad \psi_2 = \delta \psi_2,
\end{equation}
where $\delta \psi_{1,2} \ll 1$ as $x \rightarrow \pm \infty$. The corresponding asymptotic equations for the scalar fields \eqref{2-10}-\eqref{2-20} will be rewritten as:
\begin{eqnarray}
	\delta \psi_1^{\prime \prime} &=& 
	2 E_1 \delta \psi_1,
\label{psi1}\\
	\delta \psi_2^{\prime \prime} &=&  
	\left( \frac{E_1}{\lambda_1} - E_2 \right) \delta \psi_2 
\label{psi2}
\end{eqnarray}
with the exponentially fast damping solutions 
\begin{eqnarray}
	\delta \psi_1 &\approx&  C_1 
	\exp \left(- x \sqrt{2 E_1} \right),
\label{sol1} \\
	\delta \psi_2 &\approx&  
	C_2 \exp{\left(- x \sqrt{\frac{E_1}{\lambda_1} - E_2} \right)}
\label{sol2}
\end{eqnarray}
where $C_{1,2}$ are integration constants. 
\par 
The energy density 
\begin{equation}
\label{enerfy}
	\epsilon(x) = \frac{1}{2} \left(
		{\psi'_1}^2 + {\psi'_2}^2 
	\right) + \frac{\lambda_1}{4} \left(
		\psi_1^2 - \frac{E_1}{\lambda_1}
	\right)^2 + \frac{\lambda_2}{4} \left(
		\psi_2^2 - \frac{E_2}{\lambda_2}
	\right)^2 + \frac{1}{2} \psi_1^2 \psi_2^2 - \epsilon_0 
\end{equation}
of this pattern is presented in Fig.~\ref{planar_energy}. The constant $\epsilon_0$ is undefined but we have chosen it in such a way that $\epsilon(0) = 0$. The numerical calculations show that in this case $\epsilon(x = \pm \infty) = 0$.
\par 
This planar solution shows us that in this model a superconductor and gluon condensate may have a \emph{layered structure}.

\subsection{Cylindrical pattern}

In this case Eq's \eqref{1-20} \eqref{1-30} have the following form 
\begin{eqnarray}
	- \frac{d^2 \psi_1}{d\rho^2} - \frac{1}{\rho} \frac{d \psi_1}{d \rho}
	+ \left( 
		\psi_2^2 + \lambda_1 \psi_1^2 
		\right) \psi_1 &=& E_1 \psi_1, 
\label{3-10} \\ 
	- \frac{d^2 \psi_2}{d\rho^2} - \frac{1}{\rho} \frac{d \psi_2}{d \rho}
	+ \left( 
		\psi_1^2 + \lambda_2 \psi_2^2 
		\right) \psi_2 &=& E_2 \psi_2 
\label{3-20}
\end{eqnarray}
here we are searching for $\psi_{1,2}$ as real functions and the coordinate $\rho$ is the radial coordinate that is transversal to the cylindrical pattern.  The technique of solution of the system \eqref{3-10}-\eqref{3-20} is the same as for the \eqref{2-10}-\eqref{2-20} and is described in subsection \ref{plane}. 
\par
\begin{figure}[h]
\begin{minipage}[t]{.49\linewidth}
  \begin{center}
  \includegraphics[width=9cm]{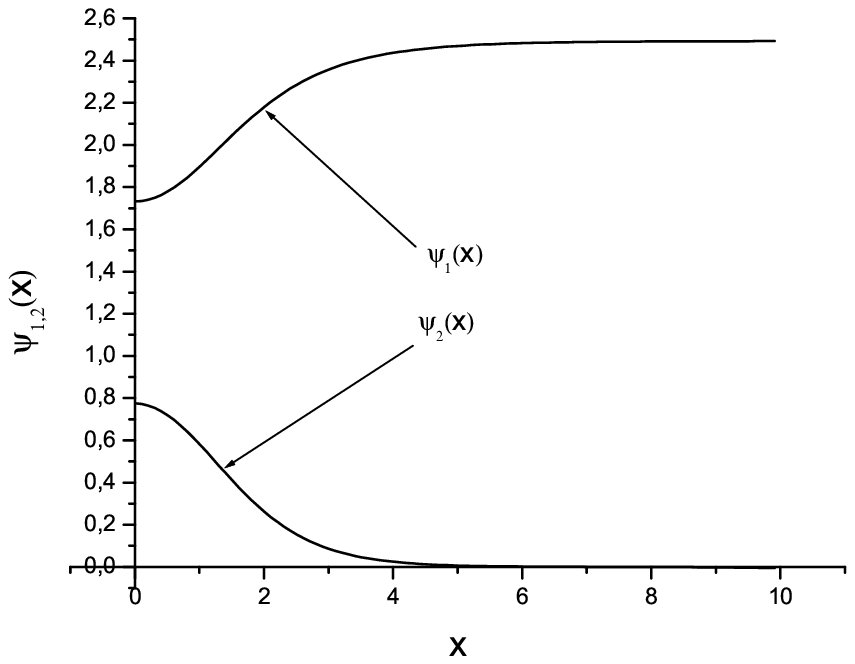}
  \caption{The wave functions $\psi_{1,2}$ for a tube, 
  $\lambda_1 = 0.05, \lambda_2 = 0.8$}
  \label{cylindr_functions}
  \end{center}
\end{minipage}\hfill
\begin{minipage}[t]{.49\linewidth}
  \begin{center}
  \includegraphics[width=9cm]{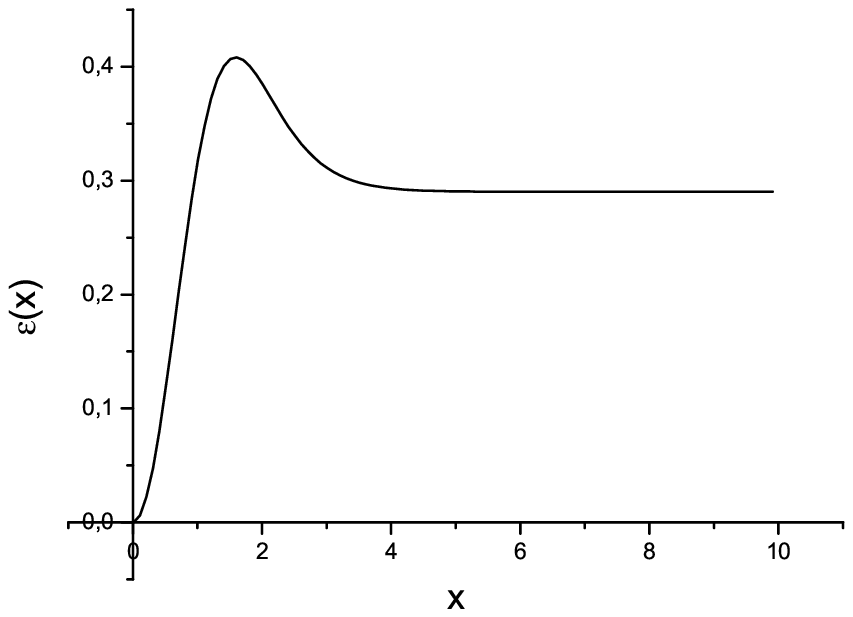}
  \caption{The energy density for a tube.}
  \label{cylindr_energy}
  \end{center}
\end{minipage}\hfill
\end{figure}

 
The boundary conditions are choosing in the following form:
\begin{alignat}{2}
\label{ini2}
	\psi_1(0)	& = \sqrt{3},				& \qquad \psi_1^\prime(0)	&=0, 
\nonumber \\
	\psi_2(0)	&=\sqrt{0.6},				& \qquad \psi_2^\prime(0)	&=0 
\end{alignat}
Then, using the above procedure for obtaining of solutions of the system  \eqref{3-10}-\eqref{3-20}, we have the results presented in Fig.~\eqref{cylindr_functions}-\eqref{cylindr_energy}. These results are obtained for the energies $E_1 \approx .310521$ and $m_2\approx 4.6365$. As one can see from Fig.~\eqref{cylindr_functions}, 
$\psi_1 \rightarrow \sqrt{E_1/\lambda_1}$ and $\psi_2 \rightarrow 0$ as 
$\rho \rightarrow \infty$. It corresponds to asymptotic transition of the solutions to
the local minimum of the potential \eqref{1-40}. 
\par 
The asymptotical consideration of equations set \eqref{3-10}-\eqref{3-20} gives us 
\begin{eqnarray}
	\psi_1 &\approx& \sqrt{\frac{E_1}{\lambda_1}} + C_1 
	\frac{\exp \left(- \rho \sqrt{2 E_1} \right)}{\sqrt{\rho}},
\label{asymp2} \\
	\psi_2 &\approx&  
	C_2 \frac{\exp{\left(- \rho \sqrt{\frac{E_1}{\lambda_1} - E_2} \right)}}{\sqrt{\rho}}
\label{asymp3}
\end{eqnarray}
where $C_{1,2}$ are integration constants. The energy density of this pattern is presented in Fig.~\ref{cylindr_energy}. The constant $\epsilon_0$ is undefined but we have chosen it in such a way that $\epsilon(0) = 0$. The numerical calculations show that in this case 
$\epsilon(\rho = \infty) > 0$.
\par 
It is useful to compare this solution with the Abrikosov-Nielsen-Olesen flux tube solution \cite{ano}: both are tubes but Nielesen-Olesen solution has a flux of the longitudinal magnetic field which is missing in the solution presented here.

\subsection{Spherical pattern}

In this case Eq's \eqref{1-20} \eqref{1-30} have the following form 
\begin{eqnarray}
	- \frac{d^2 \psi_1}{d r^2} - \frac{2}{r} \frac{d \psi_1}{d r}
	+ \left( 
		\psi_2^2 + \lambda_1 \psi_1^2 
		\right) \psi_1 &=& E_1 \psi_1, 
\label{4-10} \\ 
	- \frac{d^2 \psi_2}{d r^2} - \frac{1}{r} \frac{d \psi_2}{d r}
	+ \left( 
		\psi_1^2 + \lambda_2 \psi_2^2 
		\right) \psi_2 &=& E_2 \psi_2 
\label{4-20}
\end{eqnarray}
here we are searching for $\psi_{1,2}$ as real functions and the coordinate $r$ is the radial coordinate in the spherical coordinate system.  The technique of solution of the system \eqref{3-10}-\eqref{3-20} is the same as for the \eqref{2-10}-\eqref{2-20} and is described in subsection \ref{plane}. 
\par
\begin{figure}[h]
\begin{minipage}[t]{.49\linewidth}
  \begin{center}
  \includegraphics[width=9cm]{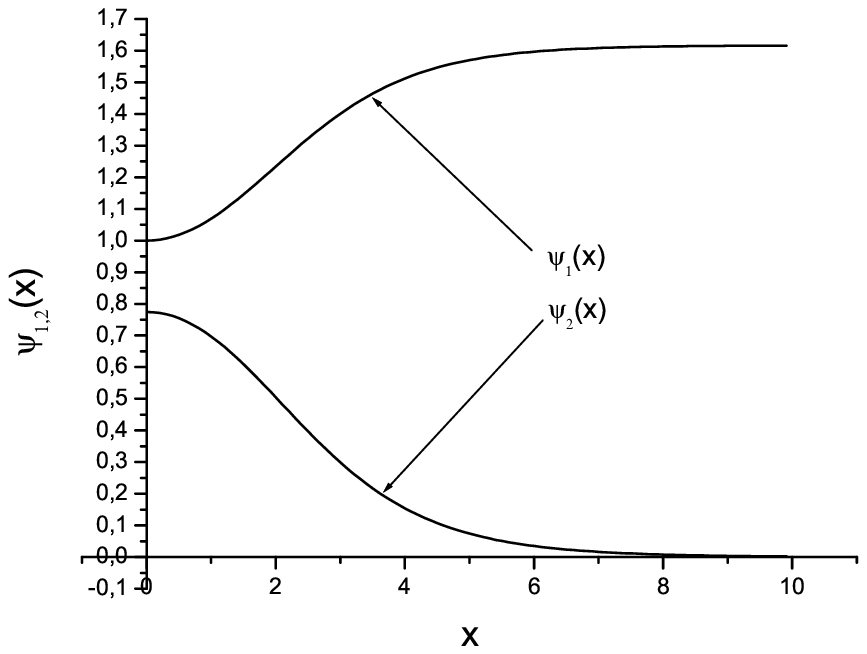}
  \caption{The wave functions $\psi_{1,2}$ for a ball, 
  $\lambda_1 = 0.1, \lambda_2 = 1$}
  \label{sphere_functions}
  \end{center}
\end{minipage}\hfill
\begin{minipage}[t]{.49\linewidth}
  \begin{center}
  \includegraphics[width=9cm]{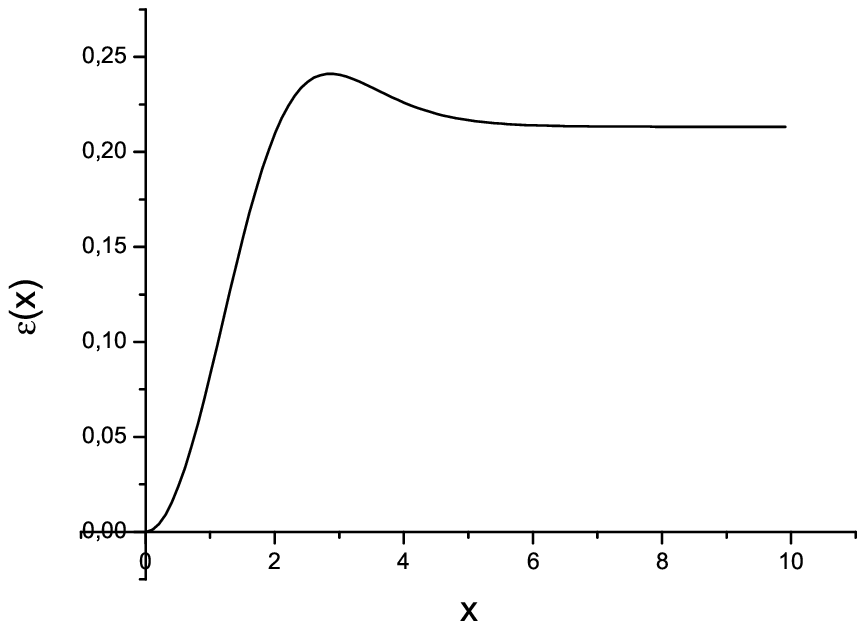}
  \caption{The energy density for a ball.}
  \label{sphere_energy}
  \end{center}
\end{minipage}\hfill
\end{figure}

 
The boundary conditions are choosing in the following form:
\begin{alignat}{2}
\label{ini3}
	\psi_1(0)	& = 1,				& \qquad \psi_1^\prime(0)	&=0, 
\nonumber \\
	\psi_2(0)	&=\sqrt{0.6},	& \qquad \psi_2^\prime(0)	&=0 
\end{alignat}
Then, using the above procedure for obtaining of solutions of the system  \eqref{3-10}-\eqref{3-20}, we have the results presented in Fig.~\eqref{cylindr_functions}-\eqref{cylindr_energy}. These results are obtained for the energies $E_1 \approx 0.261524$ and $E_2\approx 2.22826$. As one can see from Fig.~\eqref{sphere_functions}, $\psi_1 \rightarrow \sqrt{E_1/\lambda_1}$ and 
$\psi_2 \rightarrow 0$ as $\rho \rightarrow \infty$. It corresponds to asymptotic transition of the solutions to the local minimum of the potential \eqref{1-40}. 
\par 
The asymptotical consideration of equations set \eqref{4-10}-\eqref{4-20} gives us 
\begin{eqnarray}
	\psi_1 &\approx& \sqrt{\frac{E_1}{\lambda_1}} + C_1 
	\frac{\exp \left(- r \sqrt{2 E_1} \right)}{r},
\label{asymp4} \\
	\psi_2 &\approx&  
	C_2 \frac{\exp{\left(- r \sqrt{\frac{E_1}{\lambda_1} - E_2} \right)}}{r}
\label{asymp5}
\end{eqnarray}
where $C_{1,2}$ are integration constants. The energy density of this pattern is presented in Fig.~\ref{sphere_energy}. The constant $\epsilon_0$ is undefined but we have chosen it in such a way that $\epsilon(0) = 0$. The numerical calculations show that in this case 
$\epsilon(r = \infty) > 0$.

\section{Discussion}

Thus the physical consequances of the model presented above are:
\begin{itemize}
	\item a High-T$_c$ superconductor and nonperturbative gluon condensate (in quantum chromodynamics) may have the regions with different physical properties;
	\item in High-T$_c$ superconductor these regions differ from each other that in each region the interaction between Cooper electrons is carried out by different quantum (phonons, magnons, excitons and so on);
	\item in quantum chromodynamics these regions filled with different kind of a gauge condensate.
\end{itemize}
In the presented picture High-T$_c$ superconductivity and quantum chromodynamics have many common features. Both have fermions (quarks in quantum chromodynamics and electrons in High-T$_c$ superconductivity) interacting via gluons (quantum chromodynamics) and phonons, magnons and so on (in High-T$_c$ superconductivity). In quantum chromodynamics the gluons are the quanta of strongly interacting gauge field. This is the reason for a nonpeturbative quantization. The problem in quantum chromodynamics is the quark confinement which appears in the fact that two quarks interact so strongly that they can not be separated. The main opinion here is that between quarks there is a pattern (flux tube) filled with a parallel chromoelectric color field. 
\par
In the presented model of two interacting GL-equations for High-T$_c$ superconductivity the situation is similar: There are electrons which are combined into Cooper pairs (or one can say that all electrons are in a strongly correlated state) and the interactions between them carries out with phonons, magnons and so on which have a strong interaction between themselves. By analogy with quantum chromodynamics where there exist patterns filled with the gauge field (flux tubes between quarks) one can assume that in High-T$_c$ superconductivity there exist patterns filled with electrons combined into Cooper pairs via either phonons, or magnons, or and so on.
\par 
For quantum chromodynamics we have proposed a similar picture: two interacting GL-equations describing two kinds of gauge condensates: the first kind of gauge potential components are in a subalgebra and the second one are in the coset.
\par 
Then we have shown that the set of two interacting GL-equations have very interesting regular solutions not found in one GL-equation. These solutions are patterns where one kind of a  condensate is pushed out by another one. 

\begin{acknowledgments}
V.D. acknowledges D. Singleton for the invitation to do research at Fresno State University and the support of a CSU Fresno Provost Award Grant. 
\end{acknowledgments}

\end{document}